\documentclass[aps,nofootinbib,superscriptaddress, showpacs,preprintnumbers,  nofootinbibt,twocolumn]{revtex4-2}
\usepackage{epsfig}
\usepackage{multirow}
\usepackage{subcaption}
\usepackage{eurosym}
\usepackage{dcolumn}
\usepackage{bm}
\usepackage{enumerate}
\usepackage{float}
\usepackage{epstopdf}
\usepackage{amsmath}
\usepackage{bm}
\usepackage{amsfonts}
\usepackage{amssymb}
\usepackage{graphicx}
\usepackage{alphalph,mathtools}
\usepackage{etoolbox}
\usepackage{color}
\usepackage{booktabs}
\usepackage{hyperref}
\hypersetup{colorlinks,citecolor=blue}
\usepackage{footnote}
\usepackage{makecell,tabularx}

\def\be{\begin{equation}}
\def\ee{\end{equation}}
\def\bea{\begin{eqnarray}}
\def\eea{\end{eqnarray}}

\usepackage{relsize,exscale}

\begin{document}

\title{\bf The estimation of parameters of generalized cosmic Chaplygin gas and viscous modified Chaplygin gas and Accretions around Black Hole in the background of Einstein-Aether gravity }

\author{Puja Mukherjee}
\email{pmukherjee967@gmail.com} 
\affiliation{Department of
Mathematics, Indian Institute of Engineering Science and
Technology, Shibpur, Howrah-711 103, India.}
\author{Ujjal Debnath}
\email{ujjaldebnath@gmail.com} 
\affiliation{Department of
Mathematics, Indian Institute of Engineering Science and
Technology, Shibpur, Howrah-711 103, India.}
\author{Himanshu Chaudhary}
\email{himanshuch1729@gmail.com}
\affiliation{Department of Applied Mathematics, Delhi Technological University, Delhi-110042, India} 
\affiliation{Pacif Institute of Cosmology and Selfology (PICS), Sagara, Sambalpur 768224, Odisha, India}
\affiliation{Department of Mathematics, Shyamlal College, University of Delhi, Delhi-110032, India.}
\author{G. Mustafa}
\email{gmustafa3828@gmail.com}
\affiliation{Department of Physics,
Zhejiang Normal University, Jinhua 321004, People’s Republic of China,}

%%%%%%%%%%%%%%%%%%%%%%%%%%%%%%%%%%%%%%%%%%%%%%%%%%%%%%%%%%%%%%%%%%%%%%%%%%%%%%%%%%%%%%%%%%%%%%%%%%%%%%%%%%%%%%%%%%%%%%%%%%%%%%%%%%%%%%%%%%%%%%%%%%%%%%%%%%%%%%%%%%%%%%%%%%%%%%%%%%%%%%%%%%%%%%%%%%%%%%%%%%%%%%%%%%%%%%%%%%%%%%%%%%%%%%%%%%%%%%%%%%%%%%%%%%%%%%%%%%%%%%%%%%%%%%%%%%%%%%%%%%%%%%%%%%%%%%

\begin{abstract}
\noindent
In this paper, we have investigated the phenomenon of accelerated cosmic expansion in the late universe and the mass accretion process of a 4-dimensional Einstein-Aether black hole. Starting with the basics of Einstein-Aether gravity theory, we have first considered the field equations and two eminent models of Chaplygin gas, viz. generalized cosmic Chaplygin gas model and viscous modified Chaplygin gas model. Then, we obtained the energy density and Hubble parameters equations for these models in terms of some dimensionless density parameters and some unknown parameters. After finding the required parameters, we proceeded with the mass accretion process. For both models, we obtained the equation of mass in terms of the redshift function and represented the change of mass of the black hole graphically with redshift. At the same time, we have made a graphical comparison between the above-mentioned models and the $\Lambda$CDM model of the universe. Eventually, we have concluded that the mass of a 4-dimensional black hole will increase along the universe's evolution in the backdrop of Einstein-Aether gravity.
\end{abstract}
\maketitle
%\textbf{Keywords:} Accretion, Black Hole, Dark Energy, Chaplygin Gas, Einstein-Aether gravity.
\tableofcontents

\section{Introduction}\label{a}
Our universe is such an enigma, whose most of the secrets are still hidden from us. In the last century one of such concealed mystery came into light when with the help of evidences like Chandra X-ray observations \cite{allen2004constraints}, Planck observations \cite{adam2016planck}, Wilkinson Microwave Anisotropy Probe (WMAP) prediction \cite{bennett2003microwave,spergel2003first,verde20022df}, Baryon acoustic oscillation (BAO) \cite{eisenstein1998baryonic,eisenstein2005detection}, observation of Ia-type supernovae \cite{perlmutter1998cosmology,garnavich1998constraints,bahcall1999cosmic,perlmutter1999supernova,filippenko1998results}, Sloan Digital Sky Survey (SDSS) \cite{abazajian2004second,adelman2008sixth}, large-scale structure theory \cite{filippenko1998results,abazajian2004second,perlmutter1999measurements,perlmutter1998discovery,tegmark2004cosmological,perlmutter1997measurements} and observation of cosmic microwave background (CMB) which shows a peak while measuring angular power spectrum on degree scales \cite{de2000flat,lange2001cosmological,balbi2000constraints,spergel2007three} firmly indicated towards the current accelerated rate of expansion of our universe. Among all of the possibilities for such kind of accelerated expansion rate, most of the cosmologist agreed upon an exotic kind of energy named as ``dark energy" which possess quite an unique type of properties such as negative pressure and positive energy density. Dark energy has several forms, like a static form commonly known as the cosmological constant \cite{padmanabhan2003cosmological,carroll1992cosmological}. Then comes the ones that changes over time called as a scalar field or quintessence \cite{caldwell1998cosmological,peebles1988cosmology,bamba2012dark,chen2015constraints,smer2017planck,wetterich1988cosmology}. As the time progresses researchers came across many other candidates of dark energy named as phantom energy \cite{caldwell2002phantom,elizalde2004late}, k-essence \cite{armendariz2000dynamical,armendariz2001essentials}, quintom \cite{feng2005dark,guo2005cosmological}, dilaton \cite{gasperini2001quintessence}, tachyon field \cite{sen2002rolling,sen2002tachyon,gibbons2002cosmological}, hessence \cite{wei2005hessence}, DBI-essence \cite{gumjudpai2009generalized,martin2008dbi} etc. Some interacting models like holographic dark energy model \cite{li2004model,hsu2004entropy,cohen1999effective} and braneworld model of Universe \cite{sahni2003braneworld} have been investigated, besides considering the theory of quantum gravity, models like the agegraphic dark energy (ADE) model and the new agegraphic dark energy (NADE) model have been introduced \cite{cai2007dark,wei2008new}.\\\\
Chaplygin gas(CG) models are another prominent candidate of dark energy, which is basically a combination of dark matter and dark energy, constructed to measure the acceleration rate of our universe. It has different types such as; pure Chaplygin gas \cite{kamenshchik2001alternative,bilic2002unification,gorini2005chaplygin}, more general form named generalized Chaplygin gas (GCG) \cite{alam2003exploring,gorini2003can,bento2002generalized}, modified version commonly popular as modified Chaplygin gas (MCG) \cite{benaoum2012modified,debnath2004role}, which propagates between dust (or radiation) stage to $\Lambda$CDM stage. In 2003, a new model of CG, named as generalized cosmic Chaplygin gas (GCCG) model has been introduced \cite{gonzalez2003you} with some striking features like, this model can be made stable and without any unphysical behaviour, even though the vacuum fluid satisfies the condition of phantom energy. For such unique properties, this model can easily overcome the concept that the phantom energy era will finally lead us to Big-Rip as, in finite time, derivative of scale-factor with respect to time gives infinite value. The GCCG model has fewer constraints than GCG and MCG models, so it can adapt to any cosmological framework with just some appropriate choices of parameters. Thus it is more universal in nature as well as in this model black holes does not evaporate to zero masses and Big-Rip singularity can be avoided quite effortlessly. Intriguing by such flexible properties of this model several researches have been conducted on this every now and then \cite{chakraborty2007generalized,rudra2013role,bhadra2012accretion,sharif2014effects,rudra2012dynamics,sharif2014phantom,eid2018schwarzschild,sharif2013reissner}. Another addition to this list of dark energy models is viscous modified Chaplygin gas (VMCG) model, which have the property of bulk viscosity.Its cosmological modeling has been done in classical and loop quantum cosmology \cite{aberkane2017viscous}, the dynamical behaviour of VMCG has been studied in classical and loop quantum cosmology with the help of some recent observational data \cite{mebarki2019dynamical}, the evolution of VMCG in $f(R, T)$ gravity with respect to flat FLRW universe has been studied \cite{sharif2017interaction}, and it's interaction effects on f(R, ${\mathcal{G}}$, ${\mathcal{T}}$) gravity has been investigated \cite{mavoa2023effect}.\\\\
Apart from this, scientists came across another very aspiring theory to explain the current accelerated expansion phase of our universe known to be the modified gravity theory, which stands on the concept that although general relativity works fine for small scales but in cases of larger cosmological distances it needs modification. This theory gives priority to the arise of cosmic acceleration due to the dynamics of modified gravity framework. It has quite a good dynamical approach regarding the alternate of $\Lambda$CDM model of the Universe. Most common type of modified gravity is basically the DGP brane-world model of gravity \cite{dvali2000metastable}, in an alternate way we can change the law of gravity by considering the $f(R)$ theory of action in the place of usual Einstein-Hilbert action \cite{de2010f,nojiri2011unified}. Then there are some other models to explain the concept of the present time acceleration rate of our Universe \cite{clifton2005power}. This new approach opened several ways for researchers owing to which numerous researches have been carried out on this topic \cite{dvali20004d,jacobson2001gravity,abdalla2005consistent,hovrava2009membranes,barrow2012some,meng2012einstein,nojiri2005modified,tsujikawa2010modified}. There are other forms of modified gravities including $F(T)$ gravity \cite{yerzhanov2010accelerating}, $F(G)$ gravity \cite{antoniadis1994singularity}, Ho{\v{r}}ava-Lifshitz gravity \cite{hovrava2009membranes}, Brans-Dicke gravity \cite{brans1961mach}, Gauss-Bonnet gravity \cite{nojiri2005modified} etc.\\\\
Einstein-Aether theory is also among one of these modified gravity theories which has now been a very well known topic as it deals with the violation of Lorentz symmetry principle \cite{jacobson2001gravity,eling2004static,jacobson2004einstein,eling2006einstein,foster2006post}.At first  Jacobson and Mattingly developed this theory \cite{jacobson2001gravity,jacobson2004einstein} which later has been generalized by Zlosnik et al. \cite{zlosnik2007modifying,zlosnik2008growth}. It works on each point of space in preferable timelike direction \cite{eling2006black}.Nature of singularities of a vacuum black hole solution has been found in this theory \cite{satheeshkumar2023nature}.Being verified by many observational evidences \cite{yagi2014constraints,gong2018gravitational,oost2018constraints,khodadi2021einstein}, this theory has become quite suitable for further exploration regarding it's other aspects, be it cosmological perturbations \cite{li2008detecting,battye2017cosmological}, it's effects on the gravitational waves \cite{han2009gravitational,zhang2020gravitational}, analyzing a positive energy theorem \cite{garfinkle2011positive}, finding Aetherizing Lambda by using barotropic fluid as dark energy \cite{linder2009aetherizing} or some study related to very early universe \cite{gasperini1998repulsive}. Recently analytical static aether solutions and possible range for stability of constant density stars have been obtained \cite{eling2006spherical}, paving the way for researchers to study more about black holes in Einstein-Aether gravity framework. Subsequently, studies on spherically symmetric black holes in the non-reduced Einstein–Aether scenarios have been conducted \cite{konoplya2007gravitational}, charged black holes in Einstein-Maxwell-aether theory have been investigated \cite{ding2015charged}, n-dimensional charged black holes in Einstein–Aether gravity have been discussed in detail \cite{lin2019charged}, cosmic censorship in Einstein-Aether gravity and as a result the possibility of a collapse in its dynamics leading to unstable behavior around spherical solution has been analyzed \cite{meiers2016cosmic}. Barausse et al. have examined slowly rotating black holes in the Einstein–Aether gravity scenario, which shows that their solutions do not have any naked singularities \cite{barausse2016slowly}. Also, shadows of charged and slowly rotating black holes have been studied in \cite{zhu2019shadows}. Many other black hole related researches in Einstein–Aether gravity framework have been conducted on regular intervals
\cite{rayimbaev2021dynamics,chan2022thermodynamics,wang2022optical,zhang2020spherically,liu2022thin,adam2022rotating,foster2006noether,garfinkle2007numerical,tamaki2008generic,ding2016hawking,ding2017quasinormal,lin2019gravitational}.\\\\
On the other hand, in current research world, parameter constraining and observational data analysis have gain immense popularity due to it's close to reality implications. Researchers have observed the process of tallying theoretical methods with various observational databases such as cosmological supernova type Ia data, TORNY and Gold sample databases etc. measured in surveys like- GOODS ACS Treasury program, IfA Deep Survey and several other related research programs \cite{riess2004type,choudhury2005cosmological,tonry2003cosmological,barris2004twenty}.To continue this legacy, constraining of parameter space for generalized Chaplygin gas(GCG) model taking into consideration the location of CMB radiation peak \cite{bento2003generalized} and using WMAP data \cite{bento2003wmap} have been performed, placement of constraints on GCG model according to type Ia supernovae (SNIa) observations have been studied in \cite{makler2003constraints}. Several other works on observational constraining of GCG model have been performed as well \cite{zhu2004generalized,bertolami2004latest,gong2005observational,su2007observation,sen2005generalizing,lu2009observational,park2010observational}. Also, applying constraining on cosmic microwave background radiation (CMBR) spectrum the range for the value of the parameter in the case of modified Chaplygin gas model has been found in \cite{dao2005cosmic} and parameter constraining on variable Chaplygin gas (VCG) model using gold sample database of type Ia supernova has been discussed in \cite{guo2005observational}. Cosmologists have always been keen about the concept of parameter constraining on different type of Chaplygin gas models. Hence there has been generous amount of such works in the research field of cosmology \cite{liao2013observational,carneiro2014observational,kahya2014observational,dinda2014inflationary,ranjit2014constraining,zhu2015constraining,sharov2016observational,collodel2016chaplygin,salamate2018observational,salahedin2020cosmological,debnath2021gravitational,debnath2021roles,lu2008constraints,xu2012modified,bhadra2014constraining,debnath2020gravitational,sahlu2021confronting}.Moving a step ahead researchers took interest on constraining of parameters in different gravities; like Brans-Dicke gravity \cite{wu2010cosmic,li2013constraints,morganstern1973observational,torres2002quintessence,nakamura2006constraints,debnath2021constraining,thushari2010brans,cid2011constraints,freitas2012observational,farajollahi2011stability,farajollahi2013stability,tahmasebzadeh2016brans,amirhashchi2020constraining,debnath2021observational,akarsu2020weak,prasad2020constraining,biswas2014constraining}, loop quantum cosmology  \cite{chakraborty2012observational,biswas2013constraining}, Ho{\v{r}}ava-Lifshitz gravity \cite{paul2013observational,ranjit2016observational,biswas2015observational}, f(T) gravity \cite{chaudhary2023constraints}, Chern-Simons gravity \cite{debnath2015observational}, Galileon gravity \cite{ranjit2014study}, RS II braneworld gravity \cite{ranjit2013observational}, Einstein-Aether gravity \cite{debnath2014constraining} etc. Some other remarkable works on this topic has been discussed in \cite{yang1880metastable,yang2019challenging,yang2019observational}. Motivated by these EoS parameterizations, researchers have also extensively discussed and scrutinized the parametrization of the deceleration parameter, Hubble parameter, and a few other geometrical and physical parameters in the literature. A brief summary can be found in \cite{q(z)1,q(z)2,q(z)3,q(z)4,q(z)5,q(z)6,q(z)7,q(z)8,q(z)9,q(z)10,q(z)11,q(z)12,q(z)13,q(z)14,EoSNew1,EoSnew2,EoSnew3,Eosnew4,Eosnew5,Horova1,Horova2,Horova3}. As the Universe evolves, it shifts from a deceleration phase to a late-time acceleration phase. Consequently, any cosmological model must include a transition from deceleration to acceleration to explain the Universe's evolution completely. The deceleration parameter, denoted as \(q = -\frac{a\ddot{a}}{\dot{a}^2}\), where \(a(t)\) represents the customary scale factor, plays a pivotal role in determining whether the Universe is undergoing acceleration (\(q < 0\)) or deceleration (\(q > 0\)). In this paper, we are particularly interested in generalized cosmic Chaplygin gas (GCCG) and viscous modified Chaplygin gas (VMCG) models, regarding which some amazing works have already been conducted in \cite{maity2024constraining,aberkane2017viscous,ranjit2014constraining,mebarki2019dynamical,ranjit2016observational,debnath2021observational}.\\\\
Mass accretion phenomena have been quite a significant topic for researchers devoted to the studies of black holes. This concept first emerged through the work of Bondi \cite{bondi1952spherically}. Then Michel followed this up by considering some polytropic gases and formulating the equations of motion of spherical symmetric steady-state flow of matter to and fro some condensed objects like black holes, neutron stars, etc. \cite{michel1972accretion}. This accretion process gives encouragement to the concept of existence of supermassive black holes at the center of most of the galaxies \cite{moffat1997supermassive} and it can contribute to some of the energy portion of galaxies and quasars \cite{merritt2001relationship}. Accretion of phantom energy onto Schwarzschild black hole which ultimately leads to the Big Rip has been discussed in \cite{babichev2004black,babichev2005accretion}. Phantom energy onto black holes in oscillating or cyclic universe has been studied in \cite{sun2008phantom}. One by one, accretion of modified variable Chaplygin gas and viscous generalized Chaplygin gas onto black holes \cite{jamil2009evolution}, accretion of phantom energy onto a stringy magnetically charged black hole \cite{sharif2012phantom}, accretion process of modified Hayward black hole as well as it's evaporation \cite{debnath2015accretion}, dark energy accretion onto Kerr–Newman black holes  \cite{jimenez2008evolution}, accretion of generalized cosmic Chaplygin gas (GCCG) and new variable modified Chaplygin gas (NVMCG) onto Schwarzschild and Kerr–Newman black holes \cite{bhadra2012accretion}, black holes in expanding universe \cite{kim2012dark} have been studied. Besides, accretion process of black holes and wormholes have been discussed simultaneously in \cite{madrid2010accretion} and accretion of generalized cosmic Chaplygin gas and variable modified Chaplygin gas onto Morris–Thorne wormhole have been illustrated in \cite{debnath2014accretions}.\\\\
Afterward, many pieces of research on mass accretion process of higher dimensional black holes have been evolved in \cite{kim1997renormalized,ghosh2012nonstatic,john2013accretion,hendi2011charged,mukherjee2023accretion,abbas2013thermodynamics} along with there have been several remarkable works on mass accretion process of various black holes in the framework of different gravities and models of the universe \cite{nayak2012effect,dwivedee2014evolution,lima2010analytical,sharif2011phantom,martin2009dark,rodrigues2012accretion,abbas2013thermodynamics,martin2006will,abbas2014phantom}. Greatly inspired by all of these researches, we have analyzed about the generalized cosmic Chaplygin gas (GCCG) and viscous modified Chaplygin gas (VMCG) models on the backdrop of Einstein-Aether gravity. Even though we are familiar with the concept that modified gravity is an alternative to dark energy, and that is the reason for not considering any other dark energies in the case of Einstein-Aether gravity as it may itself initiate dark energy \cite{meng2012einstein,meng2012specific}. However in our work we have exempted this hypothesis and consider the above mentioned dark energies from our side. In Sec.\ref{b}, we have discussed the basic idea of Einstein-Aether gravity. In Sec.\ref{c}, we have considered the GCCG model of dark energy and obtained its energy density equation as well as the equation of the Hubble parameter associated with it in the form of some dimensionless density parameters. In Sec.\ref{d} we have done the same for the VMCG model of dark energy. In Sec.\ref{sec4}, we have conducted the data analysis for the above two discussed models and elaborated the outcomes graphically. In Sec.\ref{f}, we have studied the accretion process of a 4-dimensional Einstein-Aether black hole in the case of GCCG and VMCG models of dark energy and portrayed the changes of mass of the black hole with respect to redshift function graphically first in the case of GCCG induced universe then in the case of VMCG induced universe and lastly shown the comparison of the same graphically for GCCG filled universe, VMCG filled universe and $\Lambda CDM$ model of the universe side by side. Then, we discussed the results of data analysis and mass accretion process and explained the corresponding graphs in a detailed manner in Sec.\ref{i}. Finally we have summarized all of our findings in the Sec.\ref{j} as a conclusion of this work.\\\\

%%%%%%%%%%%%%%%%%%%%%%%%%%%%%%%%%%%%%%%%%%%%%%%%%%%%%%%%%%%%%%%%%%%%%%%%%%%%%%%%%%%%%%%%%%%%%%%%%%%%%%%%%%%%%%%%%%%%%%%%%%%%%%%%%%%%%%%%%%%%%%%%%%%%%%%%%%%%%%%%%%%%%%%%%%%%%%%%%%%%%%%%%%%%%%%%%%%%%%%%%%%%%%%%%%%%%%%%%%%%%%%%%%%%%%%%%%%%%%%%%%%%%%%%%%%%%%%%%%%%%%%%%%%%%%%%%%%%%%%%%%%%%%%%%%%%%

\section{Basics of Einstein-Aether gravity:}\label{b}
The equation for describing the action of the Einstein-Aether gravity theory coupled with the usual Einstein-Hilbert part action is given by the following form \cite{zlosnik2007modifying,meng2012einstein,debnath2014constraining}:
\begin{equation}\label{1}
S=\int{d^4x\sqrt{-g}\left[\frac{R}{16\pi G}+\mathbb{L}_V+\mathbb{L}\right]},
\end{equation}
where $\mathbb{L}_V$ is the Lagrangian density of the vector field and $\mathbb{L}$ represents the same for all of the other matter fields. The Lagrangian density related to the vector field is given by \cite{zlosnik2007modifying,meng2012einstein,debnath2014constraining}:
\begin{equation}\label{2}
\mathbb{L}_V=\frac{m^2}{16\pi G}E(K)+\frac{1}{16\pi G}\lambda(B^iB_i+1),
\end{equation}
\begin{equation}\label{3}
K=m^{-2}K^{ij}_{kl}\nabla_iB^k\nabla_jB^l ,
\end{equation}
\begin{equation}\label{4}
K^{ij}_{kl}=e_1g^{ij}g_{_{kl}}+e_2\delta^i_k\delta^j_l+e_3\delta^i_l\delta^j_k,
\end{equation}
Here, $e_i$ are dimensionless constants, and $m$ is a coupling constant with mass dimension. $\lambda$ serves as a Lagrange multiplier enforcing the unit constraint for the time-like vector field. $B^i$ represents a contravariant vector, $g_{ij}$ is the metric tensor, and $E(K)$ denotes an arbitrary function of $K$. Eqn (\ref{1}) yields the field equations \cite{debnath2014constraining}:
\begin{equation}\label{5}
G_{_{ij}}=T^V_{ij}+8\pi G T_{ij},
\end{equation}
and
\begin{equation}\label{6}
\nabla_i(E'Q^i_j)=2\lambda B_j ,
\end{equation}
Where,
\begin{align}
E'=\frac{dE}{dK}~~and~~Q^i_j=2K^{il}_{jk}\nabla_lB^k
\end{align}
Where, $T_{ij}$ is given by \cite{meng2012einstein,debnath2014constraining}:
\begin{equation}\label{7}
T_{_{ij}}=(\rho+p)\mathfrak{u}_i\mathfrak{u}_j+pg_{_{ij}} ,
\end{equation}
which gives us the expression of energy-momentum tensor for all matter fluids. Here, $p$ and $\rho$ stand for all matter fluids' respective pressure and density. Also, $\mathfrak{u}_i=(1,0,0,0)$ denotes the vector of fluid 4-velocity. Again, $T^V_{ij}$ stands for the energy-momentum tensor of vector field whose expression is given as follows \cite{meng2012einstein,debnath2014constraining}:
\begin{equation}\label{8}
\begin{split}
T_{ij}^{V} = & \frac{1}{2}~\nabla_{l}\left[\left({Q_{(i}}^{l}B_{j)}-{Q^{l}}_{(i}B_{j)}-Q_{(ij)}B^{l}\right)E'\right]\\
&-C_{(ij)}E'+\frac{1}{2}~g_{ij}m^{2}E+\lambda B_{i}B_{j},
\end{split}
\end{equation}
together with
\begin{equation}\label{9}
 C_{ij}=-e_1[(\nabla_l B_i)(\nabla^l B_j)-(\nabla_i B_l)(\nabla_j B^l)] ,
\end{equation}
Here, $(ij)$ subscript shows the symmetricity property of the indices involved. Also, $B^i=(1,0,0,0)$ represents nonvanishing time-like unit vectors that follow the relation: $B^iB_i=-1$. Now, the Friedmann-Robertson-Walker (FRW) metric of the universe
given by:
\begin{equation}\label{10}
ds^{2}=-dt^{2}+a^{2}(t)\left[\frac{dr^{2}}{1-kr^{2}}+r^{2}\left(d\theta^{2}+sin^{2}\theta d\phi^{2}\right) \right],
\end{equation}
Here, $a(t)$ denotes the scale factor and $k$ is called the curvature scalar, which can take values $-1, 0, 1$. Again, from the above two Eqn.(\ref{3}) and Eqn.(\ref{4}) we will have the following one \cite{debnath2014constraining}:
\begin{equation}\label{11}
K=m^{-2}\left(e_{1}g^{ij}g_{kl}+e_{2}\delta^{i}_{k}\delta^{j}_{l}+e_{3}\delta^{i}_{l}\delta^{j}_{k}\right)=\frac{3\beta H^{2}}{m^{2}},
\end{equation}
Here, $\beta$ is a constant given by the expression: $\beta=e_1+3e_2+e_3$. Also, from the above Eqn.(\ref{5}) we will have a modified version of Friedmann equations in the background of Einstein-Aether gravity as follows \cite{zlosnik2007modifying,meng2012einstein,debnath2014constraining}:
\begin{equation}\label{12}
\beta\left(-E'+\frac{E}{2K}\right)H^{2}+\left(H^{2}+\frac{k}{a^{2}}\right)=\frac{8\pi G}{3}~\rho
\end{equation}
and
\begin{equation}\label{13}
\beta\frac{d}{dt}\left(HE'\right)+\left(-2\dot{H}+\frac{2k}{a^{2}}\right)=8\pi G(\rho+p) ,
\end{equation}
Where, $H$ is the Hubble parameter, given by: $H~(\equiv\frac{\dot{a}}{a})$. It is quite clear from the above two Eqn.(\ref{11}) and Eqn.(\ref{12}) that whenever we put the L.H.S. of the first expressions of both of the Eqn.(\ref{11}) and Eqn.(\ref{12}) equals to zero, then we will get the general field equations for Einstein's gravity. Hence we can conclude that the first expression appears due to the Einstein-Aether gravity. Again, the conservation equation is given as follows:
\begin{equation}\label{14}
\dot{\rho}+3H(\rho+p)=0
\end{equation}
If we consider that the matter field of our universe is an amalgamation of dark matter and dark energy. Then, the pressure and energy density will be of the forms:
$\rho=\rho_{_{DM}}+\rho_{_{DE}}$ and $p=p_{_{DM}}+p_{_{DE}}$ respectively.
Again, separate applications of conservation equations on dark matter and dark energy will lead us to the following equations:
\begin{equation}\label{15}
\dot{\rho}_{_{DM}}+3H(\rho_{_{DM}}+p_{_{DM}})=0
\end{equation}
and
\begin{equation}\label{16}
\dot{\rho}_{_{DE}}+3H(\rho_{_{DE}}+p_{_{DE}})=0.
\end{equation}
Since dark matter has nearly no pressure so, we can take $p_{_{DM}}=0$, also by considering the redshift equation:~$(1+z)=\frac{1}{a}$, finally the Eqn.(\ref{15}) will reduces to the form, given as: $\rho_{_{DM}}=\rho_{_{DM0}}a^{-3}$ which in terms of redshift function $z$ will be of the following form:
\begin{equation}\label{17}
\rho_{_{DM}}=\rho_{_{DM0}}(1+z)^{3}.
\end{equation}
The term $\rho_{_{DM0}}$ gives the present energy density value of dark matter.

%%%%%%%%%%%%%%%%%%%%%%%%%%%%%%%%%%%%%%%%%%%%%%%%%%%%%%%%%%%%%%%%%%%%%%%%%%%%%%%%%%%%%%%%%%%%%%%%%%%%%%%%%%%%%%%%%%%%%%%%%%%%%%%%%%%%%%%%%%%%%%%%%%%

\subsection{Generalized Cosmic Chaplygin Gas (GCCG) Model:}\label{c}
Let us consider GCCG to be a candidate of dark energy in this Einstein-Aether gravity framework.The equation of state for GCCG model is given by \cite{gonzalez2003you}:
\begin{equation}\label{18}
p_{_{DE}}=-\rho_{_{DE}}^{-\alpha}\left[ \mathcal{C} +
(\rho_{_{DE}}^{1+\alpha}-\mathcal{C})^{-\omega}\right],
\end{equation}
Here, $\mathcal{C}=\frac{\mathcal{B}}{1+\omega}-1$ and $-l<\omega<0$ with $l$ being a constant having value greater than one. Also, $\alpha$ and $\mathcal{B}$ are any two constants. It is quite clear from the above Eqn.(\ref{18}) that whenever $\omega\rightarrow 0$, it will reduces to the equation of state for generalized Chaplygin gas(GCG). Again, whenever $\omega\rightarrow -1$, it will give us the de Sitter fluid. Now, using Eqn.(\ref{16}) and Eqn.(\ref{18}) together, we will have the energy density equation of GCCG as
follows \cite{chakraborty2007generalized,debnath2021observational}:
\begin{equation}\label{19}
 \rho_{_{DE}}=\left\{\mathcal{C} +
\left(1+\mathcal{D}~a^{-3(1+\alpha)(1+\omega)}\right)^{\frac{1}{1+\omega}}\right\}^{\frac{1}{1+\alpha}},
 \end{equation} 
where, $\mathcal{D}$ represents a constant. We can write the above Eqn.(\ref{19}) in the following form \cite{debnath2021observational}:
\begin{equation}\label{20}
\begin{aligned}
\rho_{_{DE}} = & \rho_{_{DE0}}\left\{\mathcal{A} +(1-\mathcal{A}) \right. \\
& \left. \left(\mathcal{A}_{s}+(1-\mathcal{A}_{s})(1+z)^{3(1+\alpha)(1+\omega)}\right)^{\frac{1}{1+\omega}} \right\}^{\frac{1}{1+\alpha}},
\end{aligned}
\end{equation}
Here, $\mathcal{A}_{s}=\frac{1}{1+\mathcal{D}}$, $\mathcal{A}=\left(1+\mathcal{C}^{-1}\mathcal{A}_{s}^{-\frac{1}{1+\omega}}\right)^{-1}$,
and $\rho_{_{DE0}}^{1+\alpha}=\mathcal{C}+\mathcal{A}_{s}^{-\frac{1}{1+\omega}}$. Also, $z$ symbolizes the redshift function and $\rho_{DE0}$ gives us the present energy density of DE.\ Now, we can choose different forms of the function $E(K)$ since it is a free function of $K$. According to
\cite{zlosnik2007modifying,zuntz2010vector}; $E(K)$ can be taken as: $E(K)=\gamma(-K)^{n}$. On the other hand in
\cite{meng2012einstein,meng2012specific}; $E(K)$ has been taken
as: $E(K)=\gamma\sqrt{-K}+\sqrt{\frac{3K}{\beta}}~\ln(-K)$. But for easier calculation we will take $E(K)$ as \cite{debnath2014constraining}:
$E(K)=(\frac{2}{\beta})K(1-\epsilon K)$, where $\epsilon$ is a constant. So, if we solve Eqn.(\ref{12}) considering the above form of $E(K)$ then we will have the expression for Hubble parameter in terms of redshift function $z$ as follows:
\begin{equation}\label{21}
\begin{aligned}
H^2(z) = & \frac{m}{3\sqrt{3\epsilon\beta}}\left[8\pi G \rho_{_{DM0}}(1+z)^{3}-3k(1+z)^{2} \right. \\
& \left. +8\pi G \rho_{_{DE0}}\left\{\mathcal{A} +(1-\mathcal{A}) \right. \right. \\
& \left. \left. \quad \left(\mathcal{A}_{s}+(1-\mathcal{A}_{s})(1+z)^{3(1+\alpha)(1+\omega)}\right)^{\frac{1}{1+\omega}} \right\}^{\frac{1}{1+\alpha}} \right]^\frac{1}{2},
\end{aligned}
\end{equation}
Again, if we define the dimensionless parameters in the following forms: $\Omega_{_{EA}}=\frac{m}{3H_{0}\sqrt{\epsilon\beta}}$~,
$\Omega_{_{k0}}=\frac{k}{H_{0}^{2}}$~,~$\Omega_{_{m0}}=\frac{8\pi
G\rho_{_{DM0}}}{3H_{0}^{2}}$~, $\Omega_{_{DE0}}=\frac{8\pi
G\rho_{_{DE0}}}{3H_{0}^{2}}$~, the above Eqn.(\ref{21}) will reduces to the form given as follows:
\begin{equation}\label{22}
\begin{aligned}
H(z) = & H_{0}\sqrt{\Omega_{_{EA}}}\left[\Omega_{_{m0}}(1+z)^{3}-\Omega_{_{k0}}(1+z)^{2} \right. \\
& \left. +\Omega_{_{DE0}}\left\{\mathcal{A} +(1-\mathcal{A}) \right. \right. \\
& \left. \left. \quad \left(\mathcal{A}_{s}+(1-\mathcal{A}_{s})(1+z)^{3(1+\alpha)(1+\omega)}\right)^{\frac{1}{1+\omega}} \right\}^{\frac{1}{1+\alpha}} \right]^\frac{1}{4},
\end{aligned}
\end{equation}
Also, the above Eqn.(\ref{22}) and Eqn.(\ref{12}) will together leads to the following relation:
\begin{equation}\label{23}
\sqrt{\Omega_{_{EA}}}~\left[\Omega_{_{m0}}+\Omega_{_{DE0}}-\Omega_{_{k0}}\right]^{\frac{1}{4}}=1.
\end{equation}
Combining the above equations, we can obtain the following:
\begin{widetext}
\begin{multline}\label{24}
H(z)=H_{0}\left[\left\{\Omega_{_{m0}}(1+z)^{3}-\Omega_{_{k0}}(1+z)^{2}\right\}\Omega_{_{EA}}^2 \right.\\
\left.\left.+\{1-\Omega_{EA}^2(\Omega_{m0}-\Omega_{k0})\right\}
\left\{\mathcal{A} +(1-\mathcal{A})
\left(\mathcal{A}_{s}+(1-\mathcal{A}_{s})(1+z)^{3(1+\alpha)(1+\omega)}\right)^{\frac{1}{1+\omega}}\right\}^{\frac{1}{1+\alpha}}\right]^\frac{1}{4}.
\end{multline}
\end{widetext}

%%%%%%%%%%%%%%%%%%%%%%%%%%%%%%%%%%%%%%%%%%%%%%%%%%%%%%%%%%%%%%%%%%%%%%%%%%%%%%%%%%%%%%%%%%%%%%%%%%%%%%%%%%%%%%%%%%%%%%%%%%%%%%%%%%%%%%%%%%%%%%%%%%%

\subsection{Viscous Modified Chaplygin Gas (VMCG) Model:}\label{d}
If we appraised viscous modified Chaplygin gas (VMCG), a generalization of modified Chaplygin gas(MCG), considering the expansion process as of the cumulation of states that are out of thermal equilibrium resulting the bulk viscosity in the Einstein-Aether gravity context.The expression of pressure for VMCG is given by \cite{aberkane2017viscous,benaoum2012modified,mebarki2019dynamical}:
\begin{equation} \label{pressure_VMCG}
    p_{_{DE}}= A\rho_{_{DE}}-\frac{B}{\rho_{_{DE}}^\xi}- 3\zeta_0 \sqrt{\rho_{_{DE}}}H ,
\end{equation}
where, $A$ and $B$ are constants.Also, $\zeta_0$ stands for the positive coefficient of bulk viscosity and $\xi$ is any positive valued constant. Substituting
$p_{_{DE}}$ from Eqn.(\ref{pressure_VMCG}) into the Eqn.(\ref{17}), we get:
\begin{equation}\label{25}
    \rho_{_{DE}}= \left(\frac{B}{1+A-\sqrt{3}\zeta_0}+ \frac{C_1}{a^{3(1+\xi)(1+A-\sqrt{3}\zeta_0)}} \right)^{\frac{1}{1+\xi}},
\end{equation}
where $C_1$ represents the constant of integration. The above expression can
be further re-written as:
\begin{equation}\label{26}
    \rho_{_{DE}}= \rho_{_{DE0}}\left\{B_s+ (1-B_s)(1+z)^{3(1+\xi)(1+A-\sqrt{3}\zeta_0)} \right\}^{\frac{1}{1+\xi}}.
\end{equation}
where $\rho_{_{DE0}}$ being the energy density value of the dark energy at the present epoch, $B_s= \frac{B}{(1+A-\sqrt{3}\zeta_0)C_1+B}$ satisfying the
conditions $0<B_s<1$ and $1+A-\sqrt{3}\zeta_0 >0$, and
$\rho_{_{DE0}}^{1+\xi}=
\frac{(1+A-\sqrt{3}\zeta_0)C_1+B}{1+A-\sqrt{3}\zeta_0}$. From the above equations, we can obtain the following:
\begin{widetext}
\begin{multline}\label{27}
H(z)=H_{0}\left[\left\{\Omega_{_{m0}}(1+z)^{3}-\Omega_{_{k0}}(1+z)^{2}\right\}\Omega_{_{EA}}^2 \right.\\
\left.+\left\{1-\Omega_{_{EA}}^2(\Omega_{_{m0}}-\Omega_{_{k0}})\right\}
\left\{B_s+ (1-B_s)(1+z)^{3(1+\xi)(1+A-\sqrt{3}\zeta_0)}
\right\}^{\frac{1}{1+\xi}}\right]^\frac{1}{4}.
\end{multline}
\end{widetext}

%%%%%%%%%%%%%%%%%%%%%%%%%%%%%%%%%%%%%%%%%%%%%%%%%%%%%%%%%%%%%%%%%%%%%%%%%%%%%%%%%%%%%%%%%%%%%%%%%%%%%%%%%%%%%%%%%%%%%%%%%%%%%%%%%%%%%%%%%%%%%%%%%%%

\section{Methodology}\label{sec4}
In cosmology, parameter estimation typically follows a Bayesian framework, where the goal is to compute the posterior distribution of parameters \(\theta\) given observed data \(D\). This is mathematically expressed as:
\begin{equation}
P(\theta \mid D) = \frac{\mathcal{L}(D \mid \theta) P(\theta)}{P(D)}
\end{equation}
Here, \(\mathcal{L}(D \mid \theta)\) represents the likelihood function, \(P(\theta)\) is the prior distribution of the parameters, and \(P(D)\) is the marginal likelihood. The posterior distribution is explored through the parameter space \(\theta\), often using algorithms such as Metropolis-Hastings \cite{calderhead2014general}. This algorithm helps a random walker navigate the parameter space, giving preference to regions with higher likelihood values. To estimate the parameters, we determine the mean and error of the distribution by analyzing the regions where the walker frequently visits and its deviations within the parameter space. By allowing the walker to take numerous steps, we can approximate the shape of the posterior distribution. For Bayesian model selection, we calculate and compare the evidence for each model. This process typically demands significant computational resources, often utilizing nested sampling methods. However, when the posteriors resemble a Gaussian distribution, simpler tools like information criteria can offer reliable model preferences \cite{bernardo2022parametric}. Our study explores new parameterizations for the dark energy equation of state (EoS) within both Einstein's gravity and Horava-Lifshitz gravity frameworks, determining the optimal values of the free parameters in each. Utilizing Markov Chain Monte Carlo (MCMC) analysis with the Polychord package facilitates efficient exploration of the parameter space and provides reliable estimates \cite{dunkley2005fast}.
\subsection{Cosmological data sets}
For our Markov Chain Monte Carlo (MCMC) simulations, we utilize a diverse set of cosmological data. This includes data from Cosmic Chronometers \cite{CC1,CC2,CC3,CC4,CC5,CC6} and Supernovae \cite{smith2020first}
\subsubsection{Cosmic chronometers (CC)}
Cosmic chronometers are valuable tools for probing cosmic expansion, providing a model-independent method for determining the Hubble constant and the expansion rate. By analyzing the age and metallicity of nearby passive galaxies, we can estimate the expansion rate \(H_{\mathrm{CC}}(z)\) at a given redshift \(z_{\mathrm{CC}}\). This estimation is based on the observation that the expansion rate can be approximated by the ratio of the redshift difference to the time difference, adjusted for the redshift: \(H_{\mathrm{CC}}(z) \approx -(\Delta z_{\mathrm{CC}} / \Delta t) /(1+z_{\mathrm{CC}})\). We collect cosmic chronometer data from various studies \cite{CC1,CC2,CC3,CC4,CC5,CC6}, covering the redshift range \(0.07 \lesssim z \lesssim 1.97\), as compiled in \cite{bernardo2022parametric}. These data provide direct constraints on the universe's expansion history. To evaluate how well theoretical predictions align with cosmic chronometer measurements at different redshifts, we calculate the chi-squared statistic \(\chi_{\mathrm{CC}}^{2}\):
\begin{equation}
\chi_{\mathrm{CC}}^{2} = \sum_{i=1}^{31} \left( \frac{H(z_{\mathrm{CC}}) - H_{\mathrm{CC}}(z_{\mathrm{CC}})}{\sigma_{\mathrm{CC}}(z_{\mathrm{CC}})} \right)^{2}
\end{equation}
In this equation, \(H_{\mathrm{CC}}(z_{\mathrm{CC}}) \pm \sigma_{\mathrm{CC}}(z_{\mathrm{CC}})\) represents the measured expansion rate at redshift \(z_{\mathrm{CC}}\) along with its associated uncertainty. We assume that the cosmic chronometer observations are uncorrelated.
\subsubsection{Type Ia supernova (SNIa)}
We utilize the Pantheon+ compilation supernovae dataset \cite{Pantheon} as our primary observational data source. This dataset encompasses 1701 data points collected from 1550 type Ia supernovae, spanning a redshift range of $0.001 \leq z \leq 2.3$. To assess the concordance between our theoretical model and the observed supernovae data, we employ the $\chi^2$ function, defined as:
\begin{equation}
\chi^2_{\text{Pantheon+}}(\Theta) = \mathbf{F}^T \cdot \mathbf{C}^{-1}_{\text{Pantheon+}} \cdot \mathbf{F}
\end{equation}
Here, $\mathbf{C}_{\text{Pantheon+}}$ denotes the covariance matrix derived from the Pantheon+ dataset, accounting for both statistical and systematic uncertainties. The vector $\mathbf{F}$ comprises discrepancies between the apparent magnitudes $m_{B,i}$ and the predicted distance moduli $\mu_{\text{model}}$, where the latter is determined by a chosen cosmological model:
\begin{equation}
\mu_{\text{model}}(z_i, \Theta) = 5 \log_{10} D_L(z_i, \Theta) + 25
\end{equation}
The luminosity distance $D_L(z_i, \Theta)$ is computed as an integral of the Hubble parameter $H(z, \Theta)$ over redshift $z$, where $\Theta$ represents the set of cosmological parameters. Additionally, we adjust the vector $\mathbf{F}$ to accommodate the influence of Cepheid hosts on the absolute magnitude $M$, resulting in the modified vector $\mathbf{F}'$:
\begin{equation}
F^\prime_i =
\begin{cases}
m_{B,i} - M - \mu_{\text{Ceph},i} & \text{if } i \text{ is in a Cepheid host} \\
m_{B,i} - M - \mu_{\text{model}}(z_i) & \text{otherwise}
\end{cases}
\end{equation}
Here, $\mu_{\text{Ceph},i}$ denotes the distance modulus associated with the Cepheid host of the $i$-th supernova, independently measured by Cepheid calibrators. Consequently, the modified $\chi^2$ function is expressed as:
\begin{equation}
\chi^2_{\text{SNIa}} = \mathbf{F}^{\prime T} \cdot \mathbf{C}^{-1}_{\text{Pantheon+}} \cdot \mathbf{F}^{\prime}
\end{equation}
%%%%%%%%%%%%%%%%%%%%%%%%%%%%%%%%%%%%%%%%%%%%%%%%%%%%%%%%%%%%%%%%%%%%
\begin{figure*}
\centering
\includegraphics[scale=0.8]{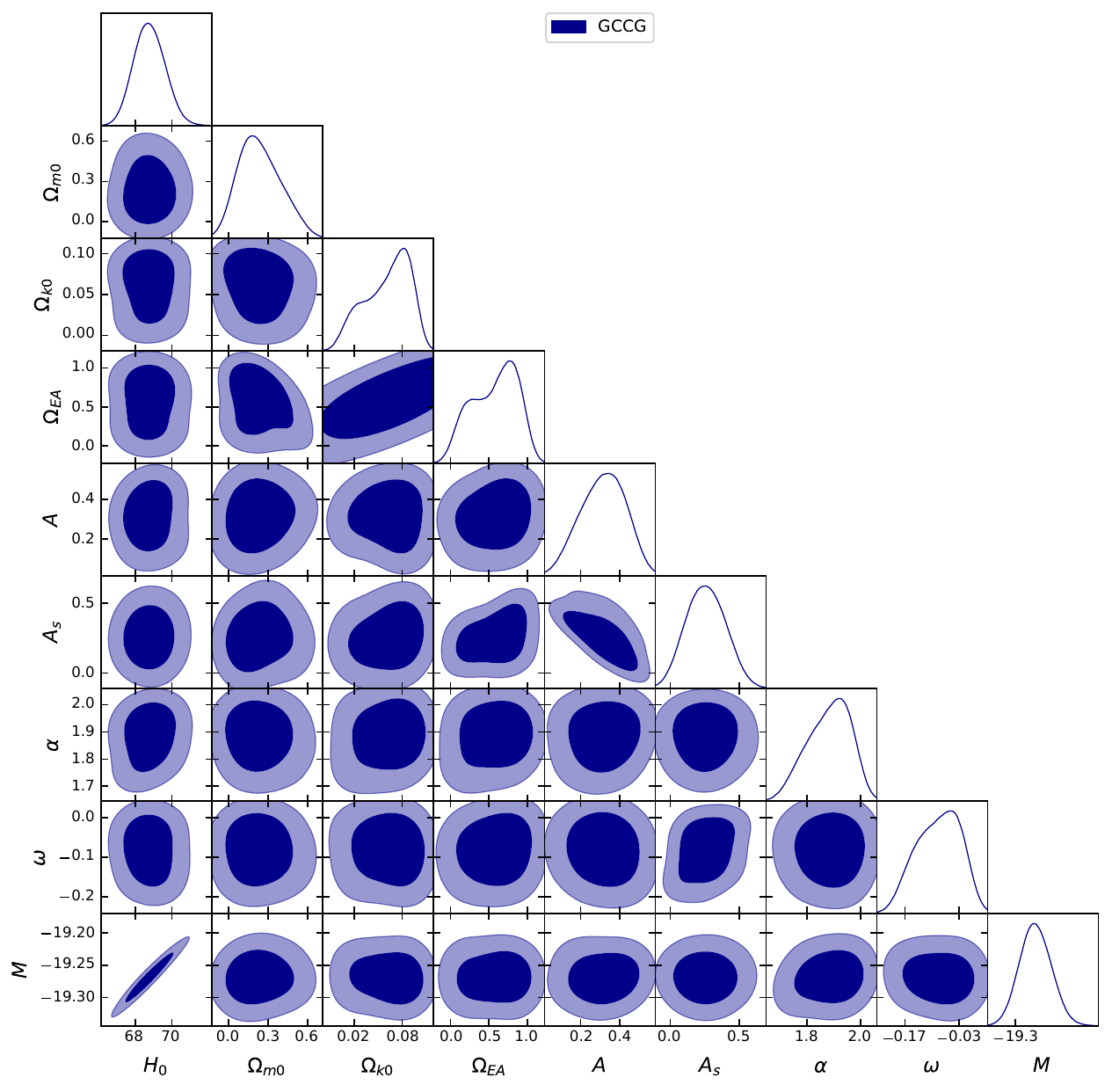}
\caption{The figure displays the posterior distribution for various observational data measurements using the Generalized Cosmic Chaplygin Gas Model, highlighting the 1$\sigma$ and 2$\sigma$ confidence levels.}\label{fig_1}
\end{figure*}
%%%%%%%%%%%%%%%%%%%%%%%%%%%%%%%%%%%%%%%%%%%%%%%%%%%%%%%%%%%%%%%%%%%%%
\begin{figure*}
\centering
\includegraphics[scale=0.8]{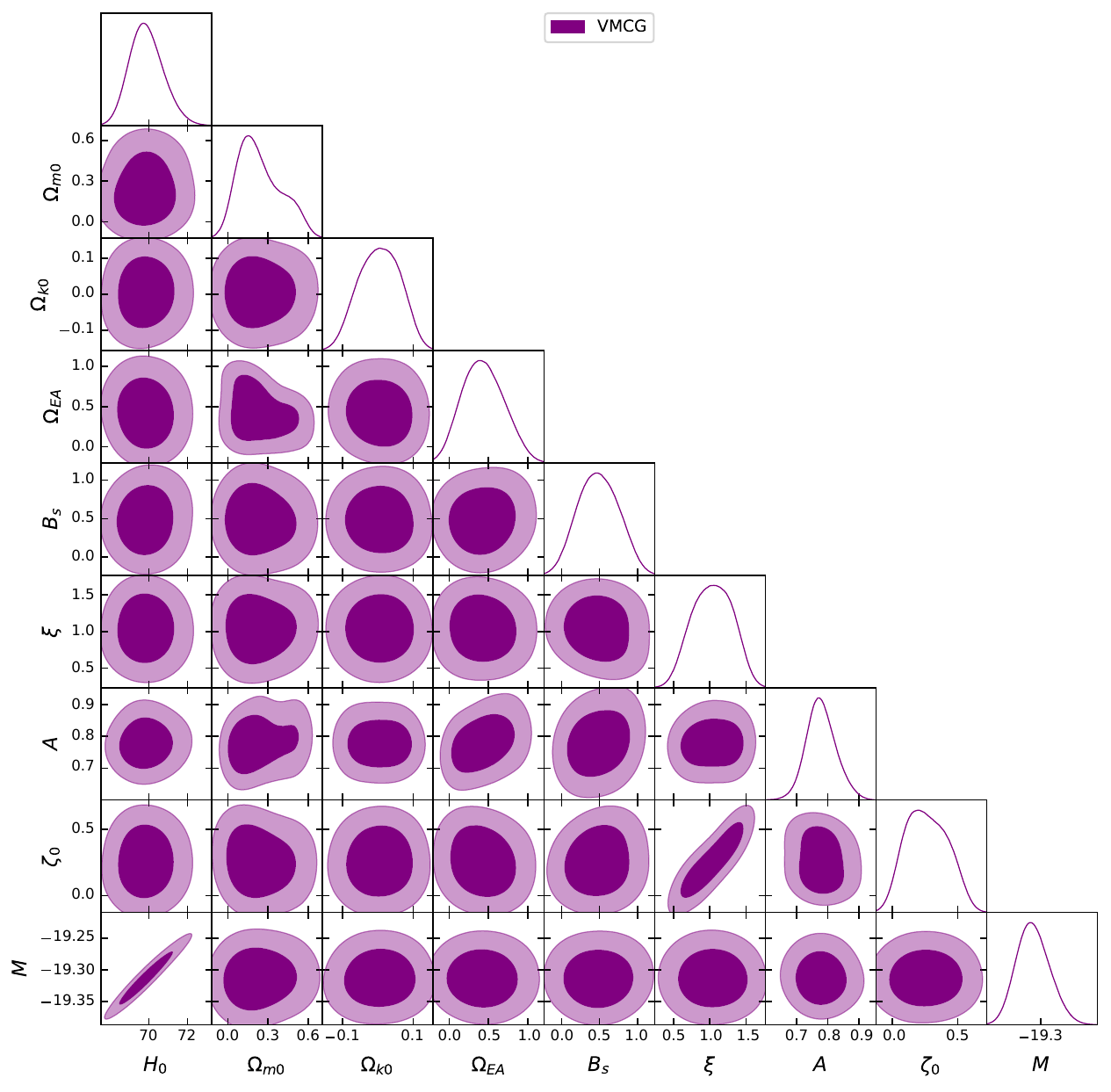}
\caption{The figure displays the posterior distribution for various observational data measurements using the Viscous Modified Chaplygin Gas Model}\label{fig_2}
\end{figure*}
%%%%%%%%%%%%%%%%%%%%%%%%%%%%%%%%%%%%%%%%%%%%%%%%%%%%%%%%%%%%%%%%%%%%%
\begin{table}
\centering
\begin{tabular}{|c|c|c|c|}
\hline
\multicolumn{4}{|c|}{MCMC Results} \\
\hline\hline
Model & Priors & Parameter & Value \\[1ex]
\hline
& $H_0$ & [50.,100.] & $69.854848_{\pm 1.259100}^{\pm 2.386935}$ \\[1ex]
$\Lambda$CDM Model &$\Omega_{m0}$ & [0.,1.] & $0.318654_{\pm 0.012822}^{\pm 0.028134}$\\[1ex]
&$\Omega_{k0}$  & [-0.1,0.1] & $-0.072537_{\pm 0.019007}^{\pm 0.025674}$ \\[1ex]
\hline
& $H_0$ & [50.,100.] & $69.765455_{\pm 0.788664}^{\pm 1.517539}$ \\[1ex]
& $\Omega_{m0}$ & [0.,0.6] & $0.295293_{\pm 0.284314}^{\pm 0.371892}$ \\[1ex]
& $\Omega_{k0}$ & [0.,0.1] & $0.061630_{\pm 0.036983}^{\pm 0.056411}$ \\[1ex]
GCCG Model & $\Omega_{EA}$ & [0.,1.] & $0.572881_{\pm 0.383366}^{\pm 0.519862}$ \\[1ex]
& $A$ & [0.1,0.5] & $0.315725_{\pm 0.128765}^{\pm 0.207903}$ \\[1ex]
& $A_{s}$ & [0.,0.6] & $0.257420_{\pm 0.135803}^{\pm 0.240137}$ \\[1ex]
& $\alpha$ & [1.7,2.] & $1.883296_{\pm 0.087537}^{\pm 0.163223}$ \\[1ex]
& $\omega$ & [-0.2,0.] & $-0.082886_{\pm 0.064216}^{\pm 0.105872}$ \\[1ex]
& $M$ & [-19.2334,0.0404] & $-19.268738_{\pm 0.023370}^{\pm 0.047949}$ \\[1ex]
\hline
& $H_0$ & [50.,100.] & $69.859683 _{\pm 0.859873}^{\pm 1.369140}$ \\[1ex]
& $\Omega_{m0}$ & [0.,0.6] & $0.298908_{\pm 0.278864}^{\pm 0.376904}$ \\[1ex]
& $\Omega_{k0}$ & [-0.1,0.1] & $0.004638_{\pm 0.064960}^{\pm 0.100714}$ \\[1ex]
VMCG Model& $\Omega_{EA}$ & [0.,1.] & $0.429626_{\pm 0.287205}^{\pm 0.401530}$ \\[1ex]
& $A$ & [0.,1.] & $0.778558_{\pm 0.043710}^{\pm 0.084330}$ \\[1ex]
& $B_s$ & [0.,1.] & $0.485455_{\pm 0.282278}^{\pm 0.458603}$ \\[1ex]
& $\xi$ & [0.5,1.5] & $1.036391_{\pm 0.322349}^{\pm 0.463270}$ \\[1ex]
& $\zeta_0$ & [0.,1.] & $0.263673_{\pm 0.181835}^{\pm 0.247089}$ \\[1ex]
& $M$ & [-19.2334,0.0404] & $-19.312621_{\pm 0.026045}^{\pm 0.039679}$ \\[1ex]
\hline
\end{tabular}
\caption{Summary of the MCMC Results.}
\label{results}
\end{table}

%%%%%%%%%%%%%%%%%%%%%%%%%%%%%%%%%%%%%%%%%%%%%%%%%%%%%%%%%%%%%%%%%%%%%%%%%%%%%%%%%%%%%%%%%%%%%%%%%%%%%%%%%%%%%%%%%%%%%%%%%%%%%%%%%%%%%%%%%%%%%%%%%%%

\section{Accretions of GCCG and VMCG around Black Hole in Einstein-Aether Gravity:}\label{f}
In this section, we are going to discuss about the mass accretion process of a 4-dimensional Einstein-Aether black hole.In this regard, let us consider a spherically symmetric and static black hole solution which have the following form:
\begin{equation}\label{28}
 ds^2=-g(r)dt^2+g^{-1}(r)dr^2+r^2(d\Theta^2+sin^2\Theta d\Phi^2).  
\end{equation}
The Einstein-Aether vector is given by \cite{wang2022optical}:
\begin{equation}\label{29}
  u^\mu(r)=\left(\delta(r)-\frac{\eta(r)}{g(r)},\eta(r),0,0\right).
\end{equation}
It would be in correspondence with the boundary condition that is the metric given by Eqn.(\ref{28}) will reduces to that of the Minkowski metric only when as $r\to+\infty$, $u^\mu=(1,0,0,0)$. Depending upon the coupling constants, we can have two types of exact black hole solutions from Einstein-Aether field of equations \cite{eling2006black,barausse2011black,berglund2013towards}. Among them the solution which is consistent with Eqn.(\ref{1}) and represents a non-rotating black hole, is commonly known as first Einstein-Aether black hole solution.In this solution the metric function $g(r)$ is given as follows \cite{ding2015charged,wang2022optical,rayimbaev2021dynamics}:
\begin{equation}\label{30}
    g(r)=1-\frac{2M}{r}-\left(\frac{27e_{_{13}}}{256(1-e_{_{13}})}\right)\left(\frac{2M}{r}\right)^4,
\end{equation}
Where, $e_{_{13}}=e_1+e_3$ and the coupling constants satisfies the following conditions: $e_{_{14}}\equiv e_1+e_4=0$, $e_{_{123}}\equiv e_1+e_2+e_3\neq 0$. Also, the functions $\delta(r)$ and $\eta(r)$ of the Einstein-Aether framework are given by \cite{berglund2013towards,wang2022optical}:
\begin{equation}\label{31}
\delta(r)=\left(\frac{3\sqrt{3}}{16\sqrt{1-e_{_{13}}}}\left(\frac{2M}{r}\right)^2+\sqrt{1-\frac{2M}{r}+\frac{27}{256}\left(\frac{2M}{r}\right)^4}\right)^{-1},   
\end{equation}\\
and
\begin{equation}\label{32}
\eta(r)=-\frac{3\sqrt{3}}{16\sqrt{1-e_{_{13}}}}\left(\frac{2M}{r}\right)^2.    
\end{equation}\\
We will get the Schwarzschild black hole solution whenever $e_{_{13}}\to 0$. Now, to investigate the accretion process of the above type of black hole in Einstein-Aether gravity background, we need to assume the energy-momentum tensor for the dark energy in the form of a perfect fluid whose equation of state is $p=p(\rho)$ given by:
\begin{equation}\label{33}
 T_{_{ij}}=(\rho+p)\mathfrak{u}_i\mathfrak{u}_j+pg_{_{ij}}   
\end{equation}
where, $\rho$ and $p$ are as usual the energy density and pressure of the dark energy.The fluid four-velocity vector taken as $\mathfrak{u}^i=\frac{dx^i}{ds}$ has the form $(\mathfrak{u}^0,\mathfrak{u}^1,0,0)$ and also satisfies the equation $\mathfrak{u}^i\mathfrak{u}_i=-1$, here only $\mathfrak{u}^0$ and $\mathfrak{u}^1$ are the two non-zero terms for this vector.Also, $\mathfrak{u}<0$ for the fluid's inward flow towards the black hole. Considering the fact that the spherical symmetry of the black hole does not have any kind of changes due to the in-falling of dark energy fluid, we are going to advance in this accretion process.
The time component of the relativistic Bernoulli's equation for the conservation of $T_{ij}$ leads us to the following:
\begin{equation}\label{34}
\mathfrak{u}r^2M^{-2}\left(\rho+p\right)\left(\sqrt{g(r)+\mathfrak{u}^2}\right)=I_0,  
\end{equation}
This is the first integral of motion, where $I_0$ stands for an integration constant having the same dimension as of the energy density. Again to obtain the second integration of motion, mostly known as the energy flux equation, we need to project $T_{ij}$ onto $\mathfrak{u}^i=\frac{dx^i}{ds}$, which is given by the equation $\mathfrak{u}_iT^{ij}_{;j}=0$  and can be written in simplified form as $\mathfrak{u}^i\rho_{,i}+(\rho+p)\mathfrak{u}^i_{;i}=0$ will lead us to the following:
\begin{equation}\label{35}
 \mathfrak{u}r^2M^{-2}exp\left[\int_{\rho_\infty}^{\rho}{\frac{d\tilde\rho}{\tilde\rho+p(\tilde\rho)}}\right]=-I_1,   
\end{equation}
where $I_1$ is another integration constant, predominantly related to the energy flux onto the black hole, given by \cite{babichev2004black,babichev2005accretion,babichev2013black}. The related negative sign has been used for the convenience of the calculation. Energy densities of the black hole at horizon and infinity are given by $\rho$ and $\rho_\infty$, respectively. Now, by dividing the above two equations, we will have the following equation:
\begin{equation}\label{36}
\left(\rho+p\right)\left(\sqrt{g(r)+\mathfrak{u}^2}\right)exp\left[-\int_{\rho_\infty}^{\rho}{\frac{d\tilde\rho}{\tilde\rho+p(\tilde\rho)}}\right]=I_2,    
\end{equation}
Here, the constant $I_2$, used in the RHS of the above equation, is given by: $I_2=-\frac{I_0}{I_1}=\rho_\infty+p(\rho_\infty)$. To compute the rate of change of mass of the black hole, i.e. $\Dot{M}$ we need to integrate the flux of the dark energy over the 4-dimensional volume of the black hole \cite{john2013accretion,debnath2015accretions}:
\begin{equation}\label{37}
 \Dot{M}=-4\pi I_0 M^2,  
\end{equation}
Now, considering the value of $I_0$, the above Eqn.(\ref{37}) will become:
\begin{equation}\label{38}
 \Dot{M}=4\pi I_1 M^2 \left[\rho_\infty+p(\rho_\infty)\right],  
\end{equation}
From the Eqn.(\ref{38}), we will get the rate of change of mass of a 4-dimensional black hole. Interestingly, for all the values of $\rho$ and $p$ which are in contrast with dominant energy condition, the above equation performs quite efficiently. So, in general, the above equation can be written as follows:
\begin{equation}\label{39}
 \Dot{M}=4\pi I_1 M^2 \left(\rho+p\right).    
\end{equation}
The above equation clearly implies that the changes in the mass of the black hole solely depends upon the term $\rho+p$.~So, in the case of phantom energy dominance, universe will contract as $(\rho+p)<0$ in this case. On the other hand in the case of quintessence kind of dark energy dominance, where $(\rho+p)>0$ the universe will expand. Again, envisaging the Einstein-Aether gravity framework, we will get the following equation by using the Eqn.(\ref{12}) and Eqn.(\ref{14}) as follows:
\begin{equation}\label{40}
 \left(\rho+p\right)=-\frac{\dot{\rho}}{3}\left[\frac{{\beta(-E'+\frac{E}{2K}+1)}}{\frac{8\pi G}{3}\rho-\frac{k}{a^2}}\right]^\frac{1}{2}. 
\end{equation}
Now, making use of Eqn.(\ref{40}) into the Eqn.(\ref{39}) and taking integration on both sides, we will obtain 
the required equation of mass in the following form:
\begin{equation}\label{41}
   M=\frac{M_0}{1+\frac{4\pi I_1 M_0}{3}\mathop{\mathlarger{\mathlarger{\int}_{\rho_0}^{\rho}}}\frac{d\rho}{\sqrt{\frac{\left(\frac{8\pi G\rho}{3}-\frac{k}{a^2}\right)}{\beta\left(1-E'+\frac{E}{2K}\right)}}}}
\end{equation}
where $M_0$ represents the current mass of the black hole and $\rho_0$ represents the current energy density which can be written as $\rho_0=\rho_{_{DM0}}+\rho_{_{DE0}}$.\\[1mm]   
The above equation can be written in a more simplified manner as follows:
\begin{widetext}
\begin{equation}\label{42}
 M=\frac{M_0}{1+\frac{4\pi I_1 M_0}{3}\mathop{\mathlarger{\mathlarger{\int}_{\rho_0}^{\rho}}}\frac{d\rho}{\sqrt{\frac{\left(\frac{8\pi G\rho}{3}-\frac{k}{a^2}\right)}{\left(\beta+\frac{15\epsilon\beta H^2}{m^2}-1\right)}}}}.   
\end{equation}
\end{widetext}
The above Eqn.(\ref{42}) is our desired equation of mass of a 4-dimensional black hole in Einstein-Aether gravity framework. Now, let us consider some Chaplygin gas models and analyzed their effects onto the mass of the Black hole during the evolution of the Universe.
\subsection{Generalized Cosmic Chaplygin Gas (GCCG) Model:}\label{g}
\begin{figure*}
    \centering
    \includegraphics[width=0.4\linewidth]{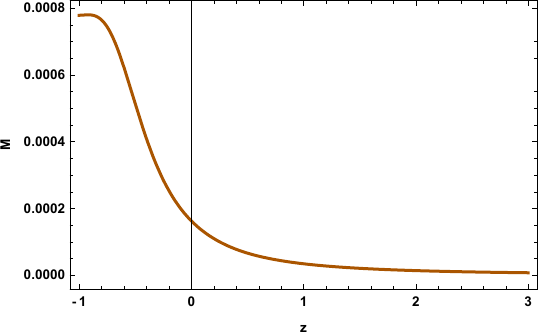}
    \caption{The changes in BH mass $M$ against redshift function $z$ on the account of GCCG accretion.}
\label{fig:my_label1}
\end{figure*}

We have already discussed this particular kind of Chaplygin gas in Sec.\ref{c}. Now, here we are going to investigate the accretion process of GCCG onto a 4-dimensional Einstein-Aether black hole. In this regard, considering Eqn.(\ref{17}), Eqn.(\ref{20}), Eqn.(\ref{23}) and Eqn.(\ref{24}) into Eqn.(\ref{42}) all at once, we will get the desired equation of mass as follows:
\fontsize{6pt}{8pt}\selectfont
\begin{widetext}
\begin{equation}\label{43}
 M=\frac{M_0}{1+\frac{4\pi I_1 M_0}{3}\mathop{\mathlarger{\mathlarger{\int}_{\rho_0}^{\rho}}}\frac{d\left(\frac{3H^2_0}{8\pi
G}\left[\Omega_{_{m0}}(1+z)^{3}+\left(\Omega_{_{EA}}^{-2}+\Omega_{_{k0}}-\Omega_{_{m0}}\right)\left\{\mathcal{A} +(1-\mathcal{A})
\left(\mathcal{A}_{s}+(1-\mathcal{A}_{s})(1+z)^{3(1+\alpha)(1+\omega)}\right)^{\frac{1}{1+\omega}}\right\}^{\frac{1}{1+\alpha}}\right]\right)}{\sqrt{\frac{\left(\frac{8\pi G\left(\frac{3H^2_0}{8\pi
G}\left[\Omega_{_{m0}}(1+z)^{3}+\left(\Omega_{_{EA}}^{-2}+\Omega_{_{k0}}-\Omega_{_{m0}}\right)\left\{\mathcal{A} +(1-\mathcal{A})
\left(\mathcal{A}_{s}+(1-\mathcal{A}_{s})(1+z)^{3(1+\alpha)(1+\omega)}\right)^{\frac{1}{1+\omega}}\right\}^{\frac{1}{1+\alpha}}\right]\right)}{3}-\frac{k}{a^2}\right)}{\left(\beta+\frac{15\epsilon\beta\left(H_0^2\left[\left\{\Omega_{_{m0}}(1+z)^{3}-\Omega_{_{k0}}(1+z)^{2}\right\}\Omega_{_{EA}}^2+\left\{1-\Omega_{_{EA}}^2(\Omega_{_{m0}}-\Omega_{_{k0}})\right\}
\left\{\mathcal{A} +(1-\mathcal{A})
\left(\mathcal{A}_{s}+(1-\mathcal{A}_{s})(1+z)^{3(1+\alpha)(1+\omega)}\right)^{\frac{1}{1+\omega}}\right\}^{\frac{1}{1+\alpha}}\right]^\frac{1}{2}
\right)}{m^2}-1\right)}}}}.   
\end{equation}   
\end{widetext}
\fontsize{10pt}{12pt}\selectfont
The above Eqn.(\ref{43}) stands for the equation of mass of a 4-dimensional Einstein-Aether black hole due to the accretion of GCCG. The effects of GCCG accretion onto the black hole mass $M$ have been portrayed by the Fig.\ref{fig:my_label1} in terms of redshift function $z$, which clearly shows the increase in black hole mass with the evolution of our universe caused by the accretion of GCCG.
\subsection{Viscous Modified Chaplygin Gas (VMCG) Model:}\label{h}
\begin{figure*}
    \centering
    \includegraphics[width=0.4\linewidth]{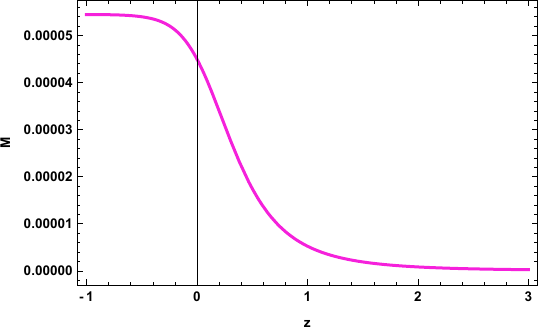}
    \caption{The changes in BH mass $M$ against redshift function $z$ on the account of VMCG accretion.}
\label{fig:my_label2}
\end{figure*}

\begin{figure*}
\begin{subfigure}{0.32\textwidth}
\includegraphics[width=\linewidth]{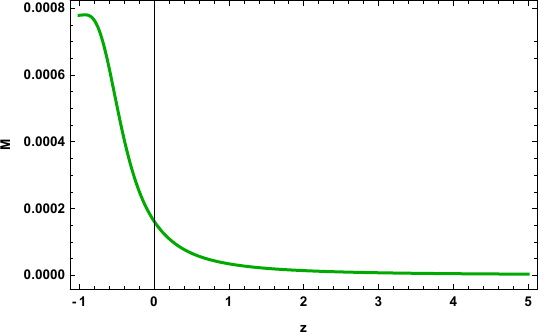}
    \subcaption{$M$ vs $z$ in GCCG Model}
    \label{fig:f 3}
\end{subfigure}
\hfill
\begin{subfigure}{0.32\textwidth}
\includegraphics[width=\linewidth]{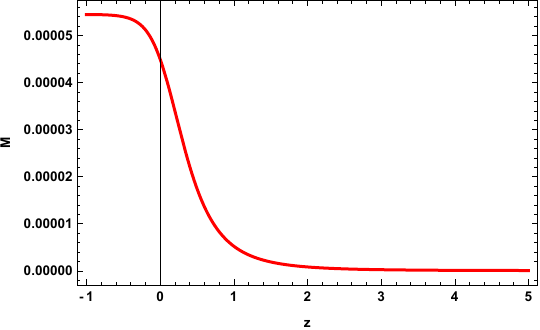}
    \subcaption{$M$ vs $z$ in VMCG Model}
    \label{fig:f 3}
\end{subfigure}
\hfill
\begin{subfigure}{0.32\textwidth}
\includegraphics[width=\linewidth]{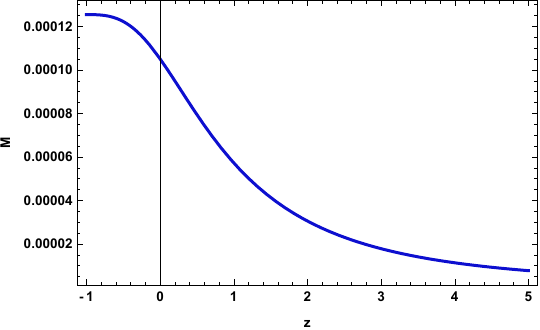}
    \subcaption{$M$ vs $z$ in $\Lambda$CDM Model}
     \label{fig:f 3}
\end{subfigure}
\caption{The changes in BH mass $M$ with redshift function $z$ for GCCG model, VMCG model and $\Lambda$CDM model of the universe in Einstein-Aether gravity framework.}
\label{fig:my_label3} 
\end{figure*}

In Sec.\ref{d} we have briefly talked about VMCG model, hence here we are going to explore its accretion process onto a 4-dimension Einstein-Aether black hole. If we consider Eqn.(\ref{17}), Eqn.(\ref{23}), Eqn.(\ref{26}) and Eqn.(\ref{27}) altogether in Eqn.(\ref{42}), we will have the desired equation of mass as follows:
\begin{widetext}
\begin{equation}\label{44}
 M=\frac{M_0}{1+\frac{4\pi I_1 M_0}{3}\mathop{\mathlarger{\mathlarger{\int}_{\rho_0}^{\rho}}}\frac{d\left(\frac{3H^2_0}{8\pi
G}\left[\Omega_{_{m0}}(1+z)^{3}+\left(\Omega_{_{EA}}^{-2}+\Omega_{_{k0}}-\Omega_{_{m0}}\right)\left\{B_s+ (1-B_s)(1+z)^{3(1+\xi)\left(1+A-\sqrt{3}\zeta_0\right)} \right\}^{\frac{1}{1+\xi}}\right]\right)}{\sqrt{\frac{\left(\frac{8\pi G\left(\frac{3H^2_0}{8\pi
G}\left[\Omega_{_{m0}}(1+z)^{3}+\left(\Omega_{_{EA}}^{-2}+\Omega_{_{k0}}-\Omega_{_{m0}}\right)\left\{B_s+ (1-B_s)(1+z)^{3(1+\xi)\left(1+A-\sqrt{3}\zeta_0\right)} \right\}^{\frac{1}{1+\xi}}\right]\right)}{3}-\frac{k}{a^2}\right)}{\left(\beta+\frac{15\epsilon\beta\left(H_0^2\left[\left\{\Omega_{_{m0}}(1+z)^{3}-\Omega_{_{k0}}(1+z)^{2}\right\}\Omega_{_{EA}}^2 \right.\\
\left.+\left\{1-\Omega_{_{EA}}^2(\Omega_{_{m0}}-\Omega_{_{k0}})\right\}
\left\{B_s+ (1-B_s)(1+z)^{3(1+\xi)(1+A-\sqrt{3}\zeta_0)}
\right\}^{\frac{1}{1+\xi}}\right]^\frac{1}{2}
\right)}{m^2}-1\right)}}}}.    
\end{equation}    
\end{widetext}
The above Eqn.(\ref{44}) stands for the equation of mass of a 4-dimensional Einstein-Aether black hole due to the accretion of VMCG.\ The effects of VMCG accretion onto the black hole mass $M$ have been portrayed by the Fig.\ref{fig:my_label2} in terms of redshift function $z$ which clearly shows the increase in black hole mass with the evolution of our universe caused by the accretion of VMCG.\ In Fig.\ref{fig:my_label3} the comparison of mass change with redshift has been depicted for GCCG model, VMCG model as well as $\Lambda$CDM model of the universe alongside eachother.
%%%%%%%%%%%%%%%%%%%%%%%%%%%%%%%%%%%%%%%%%%%%%%%%%%%
\section{Results}\label{i}
Figures \ref{fig_1} and \ref{fig_2} illustrate the \(68\%\) and \(95\%\) confidence intervals for key cosmological parameters in the Generalized Cosmic Chaplygin Gas (GCCG) Model and the Viscous Modified Chaplygin Gas (VMCG) Model, respectively. Table \ref{results} summarizes the MCMC results for three cosmological models: the \(\Lambda\)CDM model, the GCCG Model, and the VMCG Model. All three models yield consistent \(H_0\) values around 69.85, aligning with the findings in \cite{1BAO}. The \(\Lambda\)CDM model and the VMCG model exhibit tighter constraints compared to the GCCG model. However, the matter density parameter (\(\Omega_{m0}\)) is lower than the values reported in \cite{1BAO} (\(\Omega_{m0}=0.315 \pm 0.007\)) for all three cases. Then we have proceeded to observe the mass accretion process, with the help of these estimated values of parameters. The outcomes have been more than satisfactory as we have obtained the mass equation of a 4-dimensional Einstein-Aether black hole in terms of some dimensionless density parameters and redshift function $z$. After plotting the results as black hole mass $M$ vs redshift $z$ graphs in Fig.\ref{fig:my_label1} and Fig.\ref{fig:my_label2}, respectively, we can distinctly observe that with time, black hole mass increases for both the generalized cosmic Chaplygin gas (GCCG) model and viscous modified Chaplygin gas (VMCG) model. In Fig.\ref{fig:my_label3}, we can observe separately the pattern of black hole mass $M$ vs redshift $z$ graphs in the cases of generalized cosmic Chaplygin gas (GCCG), viscous modified Chaplygin gas (VMCG) as well as $\Lambda$CDM model of the universe. All of the above-mentioned models show an increase in black hole mass with the expansion of the Universe, just as predicted in the case of the dark energy-induced Universe.

%%%%%%%%%%%%%%%%%%%%%%%%%%%%%%%%%%%%%%%%%%%%%%%%%%%%%%%%%%%%%%%%%%%%%%%%%%%%%%%%%%%%%%%%%%%%%%%%%%%%%%%%%%%%%%%%%%%%%%%%%%%%%%%%%%%%%%%%%%%%%%%%%%%

\section{Conclusions}\label{j}
Black Holes have always been an object of fascination for cosmologists. Several studies have been conducted on them time and again. In this study, we have investigated Einstein-Aether black holes, basically about their mass accretion process. Firstly, we have considered the non-flat FRW model of the universe, which combines dark matter and dark energy. Then in the framework of Einstein-Aether gravity, we have taken the modified Friedmann equations. Applying the conservation equation on dark matter and dark energy separately has led us to energy densities of the dark matter and dark energy, respectively. Also, for the simplicity of the calculation, we have taken $E(K)$ as a quadratic function of $K$. Afterward, we contemplated two different kinds of dark energy models: the generalized cosmic Chaplygin gas (GCCG) model and the viscous modified Chaplygin gas (VMCG) model as the candidates for dark energy. In both cases, we have calculated the energy density and Hubble parameter equation in terms of some dimensionless density parameters and some unknown parameters one by one. Next, we selected 31 uncorrelated data points from Cosmic Chronometers and 1701 data points from the Type Ia Supernova dataset to obtain the best-fit values for the model parameters using MCMC. Finally, after estimating the required parameters, we analyzed the mass accretion process of a 4-dimensional Einstein-Aether black hole. Primarily we have obtained the mass equation for both the generalized cosmic Chaplygin gas (GCCG) model and viscous modified Chaplygin gas (VMCG) model one after another. Then we have picturized the result through the black hole mass $M$ vs. redshift function $z$ graph in Fig.\ref{fig:my_label1} for the case of GCCG accretion and in Fig.\ref{fig:my_label2} for the case of VMCG accretion respectively. Both Fig.\ref{fig:my_label1} and Fig.\ref{fig:my_label2} clearly show that the black hole mass will only grow larger with time. Again, in Fig.\ref{fig:my_label3}, we have picturized the comparison of the black hole mass $M$ vs. redshift function $z$ graph for the GCCG model, VMCG model as well as $\Lambda$CDM model of the universe in the case of Einstein-Aether gravity. So, on the whole, we can conclude that with the expansion of the universe, a 4-dimensional Einstein-Aether black hole mass will increase for accretion of dark energies like generalized cosmic Chaplygin gas and viscous modified Chaplygin gas as well as in the case of $\Lambda$CDM model of the Universe.\\\\

%%%%%%%%%%%%%%%%%%%%%%%%%%%%%%%%%%%%%%%%%%%%%%%%%%%%%%%%%%%%%%%%%%%%%%%%%%%%%%%%%%%%%%%%%%%%%%%%%%%%%%%%%%%%%%%%%%%%%%%%%%%%%%%%%%%%%%%%%%%%%%%%%%%

\section{ACKNOWLEDGEMENTS} 
PM is thankful to IIEST, Shibpur, India, for providing Institute Fellowship (JRF).

%%%%%%%%%%%%%%%%%%%%%%%%%%%%%%%%%%%%%%%%%%%%%%%%%%%%%%%%%%%%%%%%%%%%%%%%%%%%%%%%%%%%%%%%%%%%%%%%%%%%%%%%%%%%%%%%%%%%%%%%%%%%%%%%%%%%%%%%%%%%%%%%%%

\bibliographystyle{elsarticle-num}
\bibliography{myib,mybao}

\end{document}